\documentclass[aps,prl,twocolumn,superscriptaddress,reprint]{revtex4-1}

\usepackage{blindtext}
\usepackage{centernot}
\usepackage{graphicx}
\usepackage{amsmath,bbold}
\usepackage{times}
\usepackage{amssymb}
\usepackage{mathrsfs}
\usepackage{chemarr}
\usepackage{color}
\usepackage{url}
\usepackage{version}
\usepackage[hidelinks]{hyperref}
\usepackage{mwe,tikz}
\usepackage[percent]{overpic}
\usepackage{bm}
\usepackage[export]{adjustbox}
\definecolor{linkcolor}{rgb}{0,0,0.6} 
\usepackage{ stmaryrd }
\usepackage{xcolor}
\usepackage{enumerate}
\usepackage{hyperref}

\newcommand{\bF}{{\bf F}}

\newcommand{\bJ}{{\bf J}}
\newcommand{\bj}{{\bf j}}
\newcommand{\be}{{\bf e}}

\newcommand{\br}{{\bf r}}
\newcommand{\bu}{{\bf u}}
\newcommand{\bv}{{\bf v}}

\newcommand{\bomega}{\boldsymbol{\omega}}

\newcommand{\mcF}{\mathcal{F}}

\newcommand{\bC}{{\bf C}}
\newcommand{\mbd}{\mathbb{d}}
\newcommand{\rmd}{\mathrm{d}}
\newcommand{\mbF}{\mathbb{F}}
\newcommand{\bmu}{\boldsymbol{\mu}}
\newcommand{\boldeta}{\boldsymbol{\eta}}
\newcommand{\bphi}{\boldsymbol{\phi}}
\newcommand{\bpsi}{\boldsymbol{\psi}}
\newcommand{\bLambda}{\boldsymbol{\Lambda}}

\usepackage{lipsum}

\usetikzlibrary{patterns}

\begin{document}

\title{Nonequilibrium Currents in Stochastic Field Theories: a Geometric Insight}

\author{J. O'Byrne}
\affiliation{Universit\'e de Paris, Laboratoire Mati\`ere et Syst\`emes Complexes (MSC), UMR 7057 CNRS, F-75205 Paris, France}

\date{\today}

\begin{abstract}
We introduce a new formalism to study nonequilibrium steady-state
currents in stochastic field theories. We show that generalizing the
exterior derivative to functional spaces allows identifying the
subspaces in which the system undergoes local rotations. In turn, this
allows predicting the counterparts in the real, physical space of
these abstract probability currents.  The results
are presented for the case of the Active Model B undergoing
motility-induced phase separation, which is known to be out of
equilibrium but whose steady-state currents have not yet been
observed, as well as for the KPZ equation. We locate and measure these
currents and show that they manifest in real space as propagating
modes localized in regions with non-vanishing gradients of the fields.
\end{abstract}

\maketitle

Statistical physics aims at describing large-scale phenomena emerging
from interacting elementary constituents, ranging from chemicals to
animals, from bacteria to traders. Except when systems satisfy
detailed balance, no general theory can be systematically applied to
study such systems.
To understand how microscopic mechanisms drive a system out of
equilibrium, physicists have been quantifying the distance to
equilibrium using diverse observables, such as the entropy
production~\cite{lebowitz1999gallavotti, kurchan1998fluctuation,
  maes1999fluctuation, hatano2001steady, seifert2005entropy},
violations of the fluctuation-dissipation
theorem~\cite{cugliandolo2011effective, perez2003origin}, or ratchet
currents~\cite{feynman2011feynman, parrondo1996criticism,
  hanggi2009artificial, magnasco1993forced}. Among those, the
stationary probability current plays an important role since its
knowledge, together with the stationary probability measure, entirely
determine the equations of motion~\cite{zia2007probability, zia2010towards,
  liverpool2018non,supp}.  For systems driven out-of-equilibrium
by external fields~\cite{bertini2015macroscopic,
  schmittmann1995statistical, eyink1996hydrodynamics} or boundary
conditions~\cite{derrida2002large, bodineau2004current}, 
probability currents directly lead to real-space currents---e.g. of
energy or mass---that can be observed and quantified easily.
In many other situations, as in active
systems~\cite{tailleur2008statistical,nardini2017entropy,fodor2016far,obyrne2022time,galajda2007wall,flenner2020active},
surface growth problems~\cite{kardar1986dynamic} or reaction-diffusion
processes~\cite{hinrichsen2000non}, probability currents live in
high-dimensional configuration spaces and have no simple
low-dimensional projection in real space, which makes their study challenging.

While probability currents are well understood for finite-dimensional
systems~\cite{baiesi2013update, dal2021fluctuation, ge2014time,
  baiesi2010nonequilibrium, freitas2020stochastic,
  kurchan2009six,kaiser2018canonical,fang2019nonequilibrium,feng2011potential,wang2015landscape,polettini2012nonequilibrium},
collective behaviors are best described at a macroscopic scale using
field theory~\cite{tauber2014critical, kardar2007statistical,
  cardy1996scaling}. The nonequilibrium nature of such
infinite-dimensional description has attracted a lot of interest
recently~\cite{bertini2015macroscopic, nardini2017entropy,
  borthne2020time, caballero2020stealth, li2021steady} but the
identification of their \textit{probability} currents remains
ellusive.
Progress has been made in specific
situations~\cite{gladrow2016broken,battle2016broken}, but a generic
framework is crucially lacking.

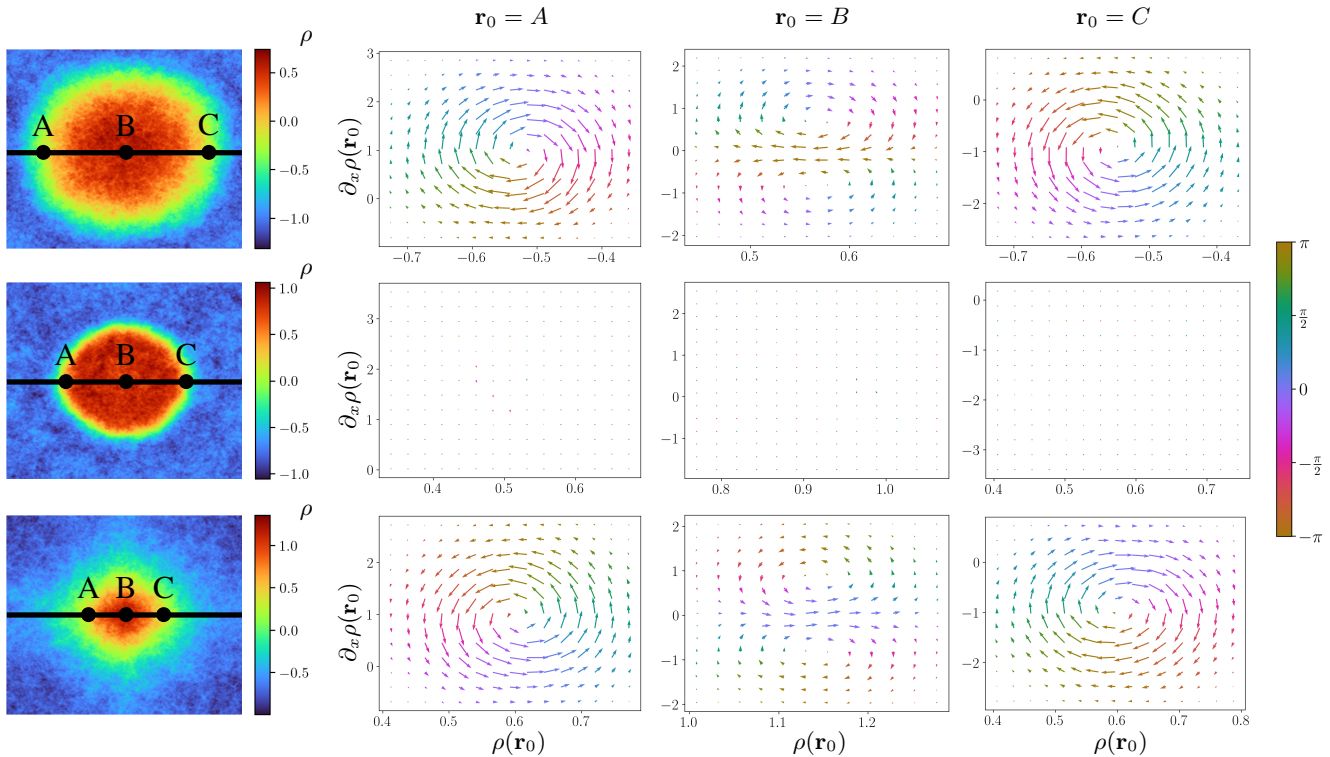
\begin{figure*}[!t]
\begin{center}
\begin{tikzpicture}
      \def\x{4.0}
      \def\xx{0}
      \def\y{-3.1}
      \def\yy{-6.2}
      \def\CurrentScale{.22}
      \def\CurrentScaleBis{.21}
      \def\DropletScale{.24}
      \def\alti{0.1}
      \def\eps{-0.6}
	
 	 \draw (-2+\eps,0+\alti) node {\includegraphics[width=\DropletScale\textwidth]{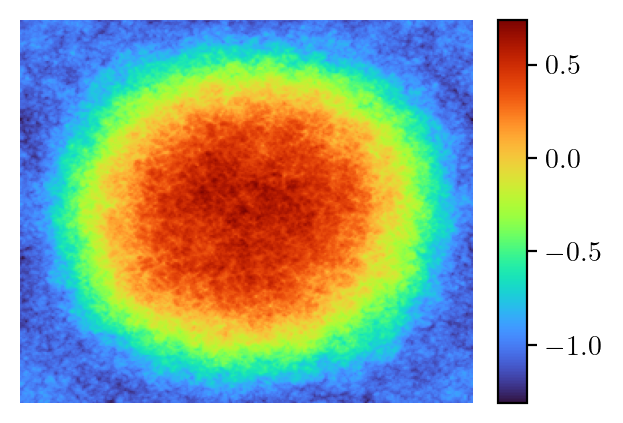}};
     \draw (\x-2,0) node{\includegraphics[width=\CurrentScaleBis\textwidth]{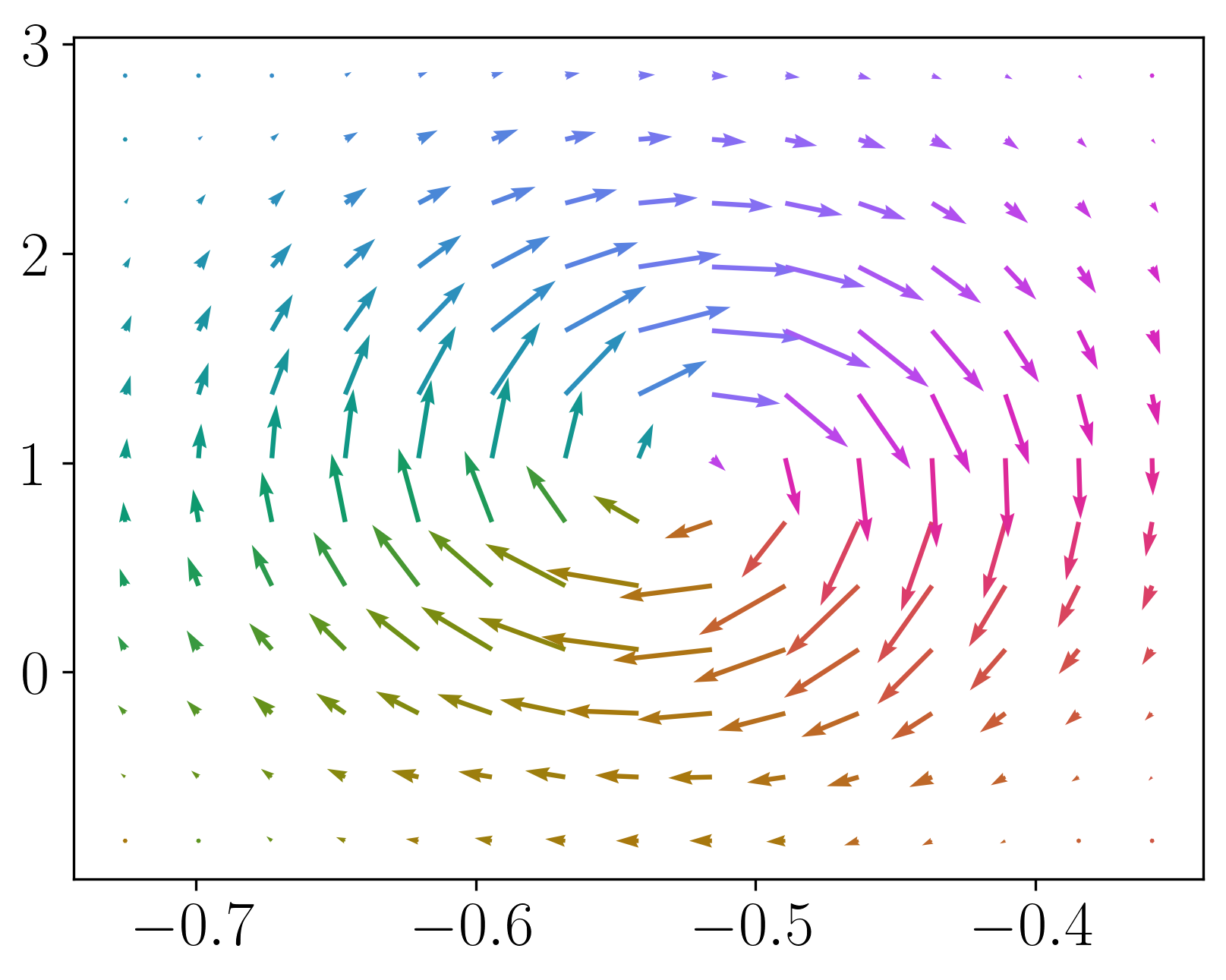}};
      \draw (2*\x-2,0) node{\includegraphics[width=\CurrentScale\textwidth]{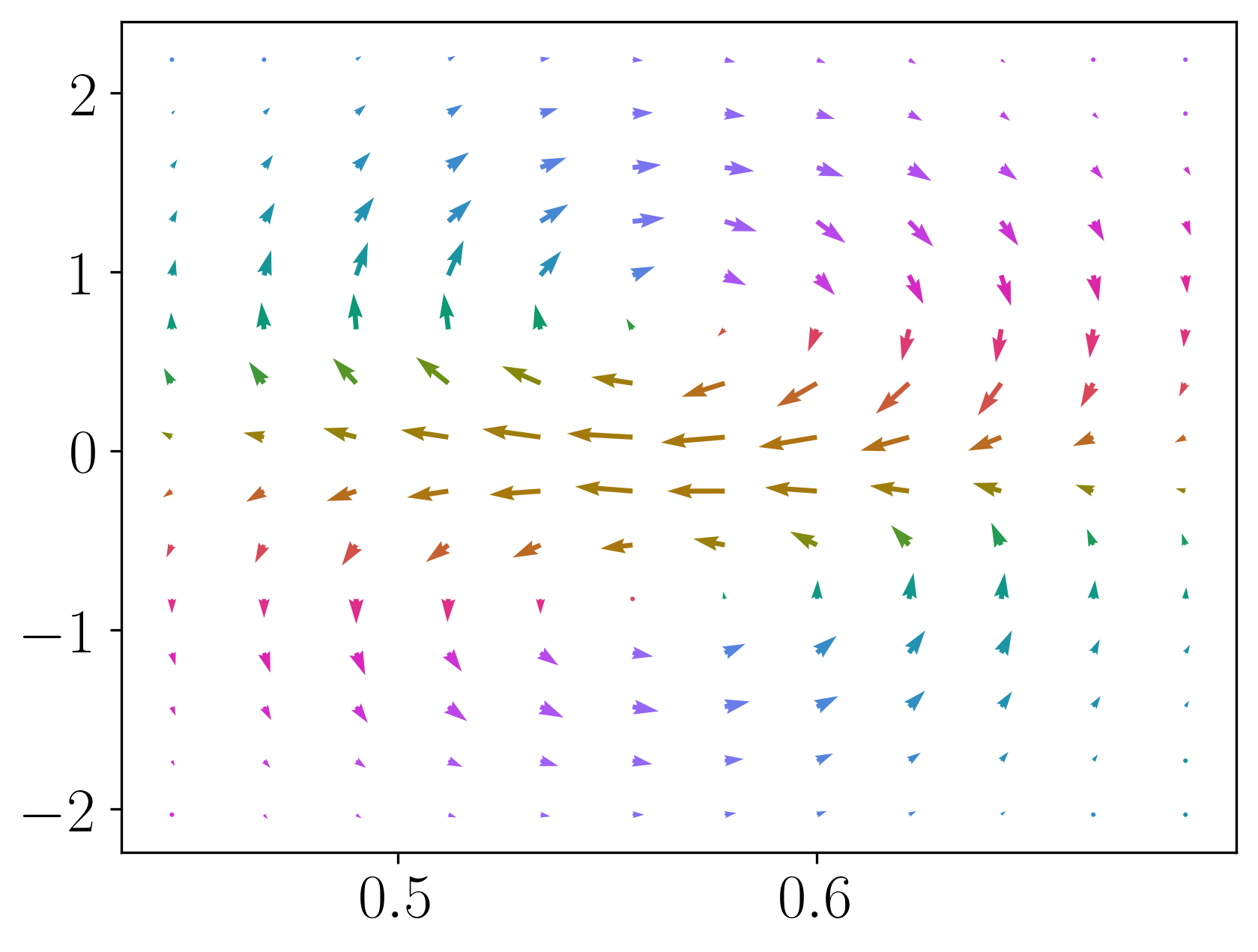}};
      \draw (3*\x-2+\xx,0) node{\includegraphics[width=\CurrentScale\textwidth]{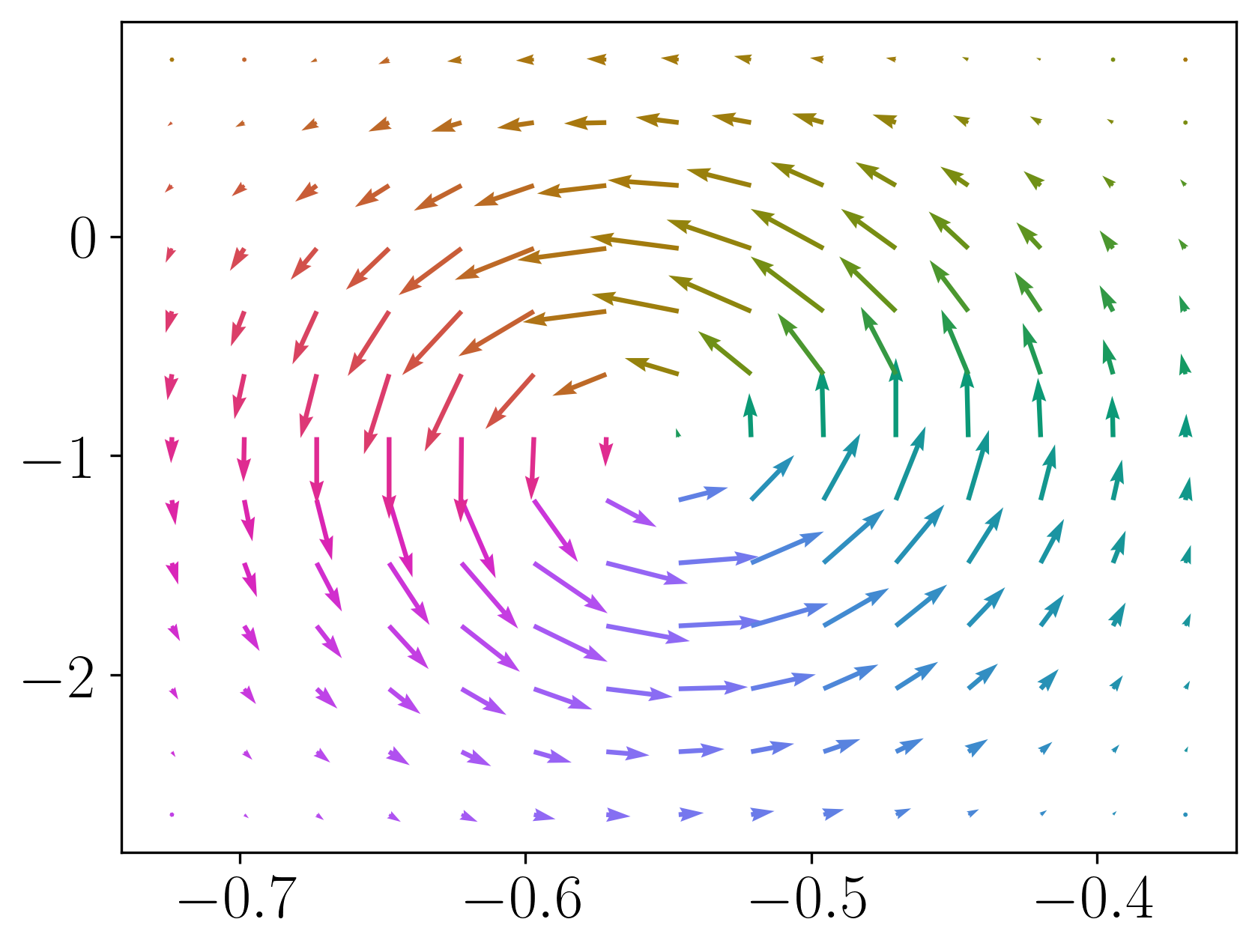}};
      \draw (-2+\eps,\y+\alti) node{\includegraphics[width=\DropletScale\textwidth]{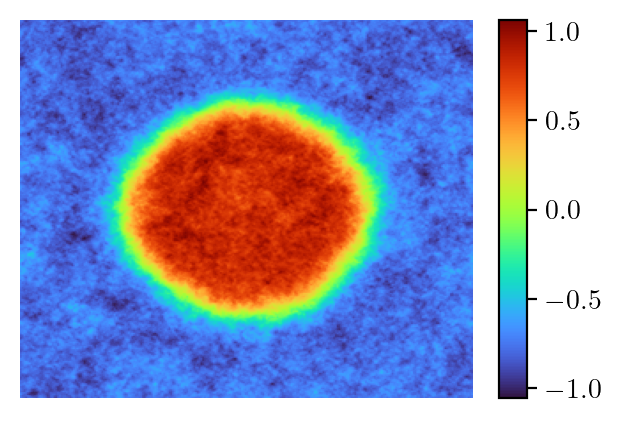}};
      \draw (\x-2,\y) node{\includegraphics[width=\CurrentScaleBis\textwidth]{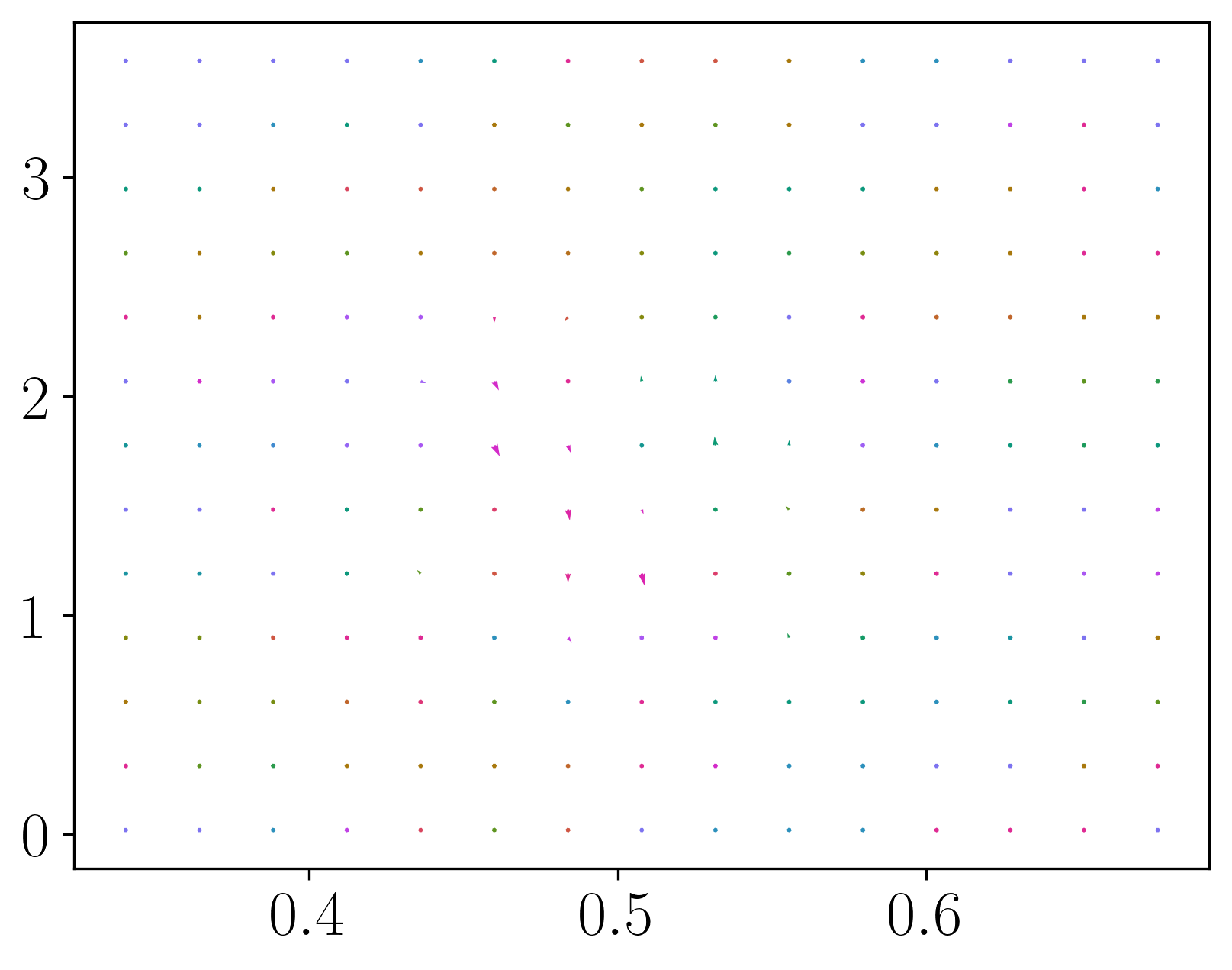}};
      \draw (2*\x-2,\y) node{\includegraphics[width=\CurrentScale\textwidth]{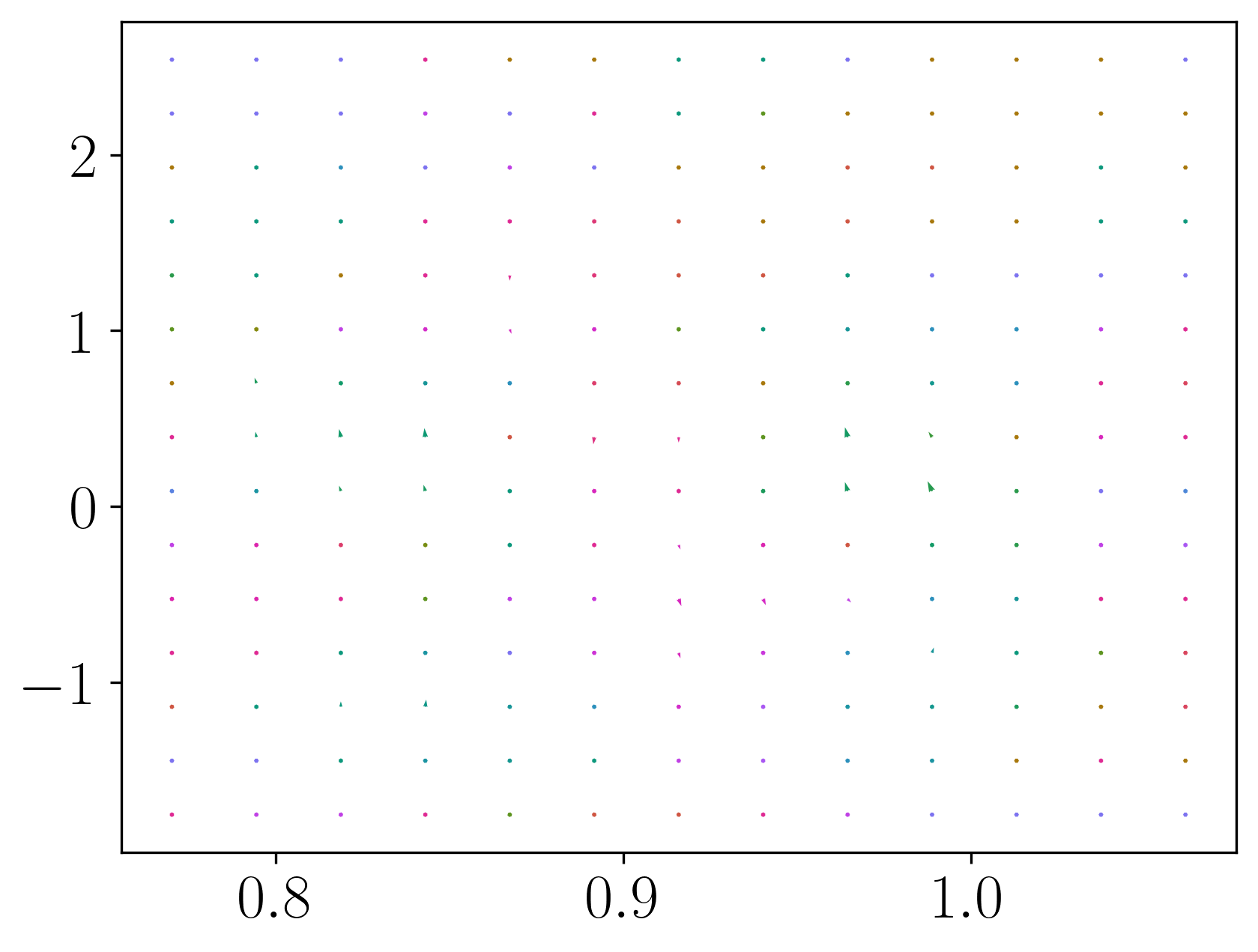}};
      \draw (3*\x-2+\xx,\y) node{\includegraphics[width=\CurrentScale\textwidth]{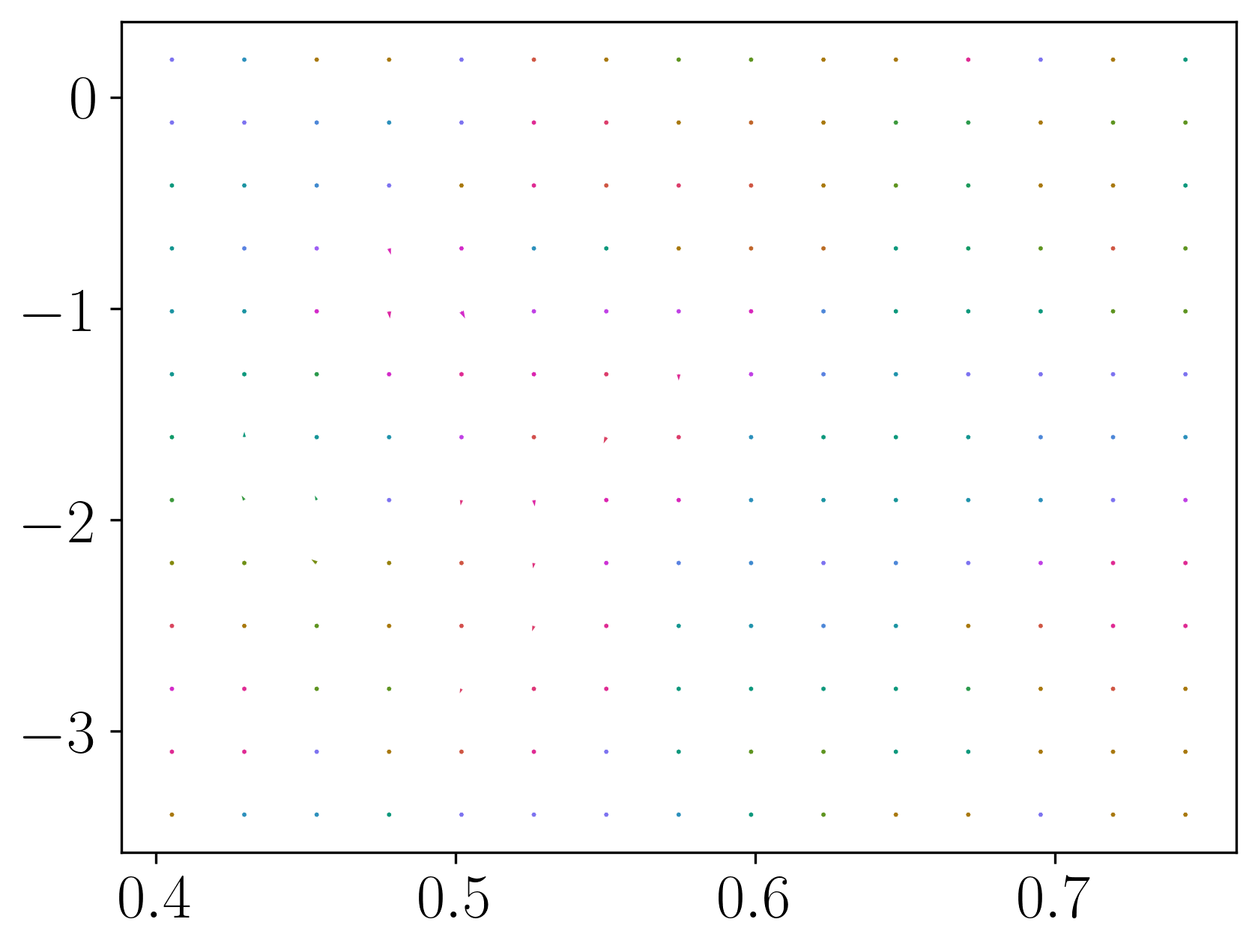}};
      \draw (-2+\eps,\yy+\alti) node{\includegraphics[width=\DropletScale\textwidth]{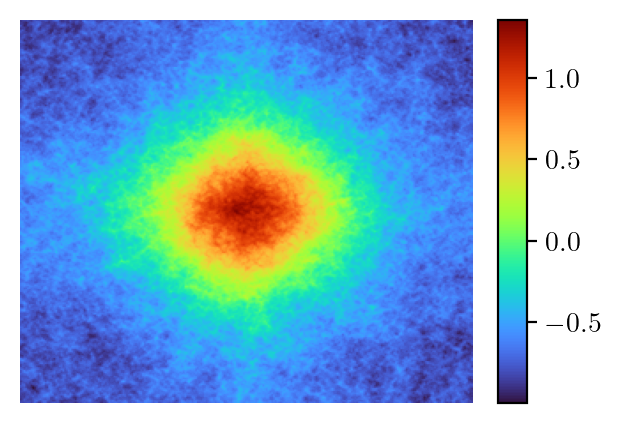}};
      \draw (\x-2,\yy) node{\includegraphics[width=\CurrentScaleBis\textwidth]{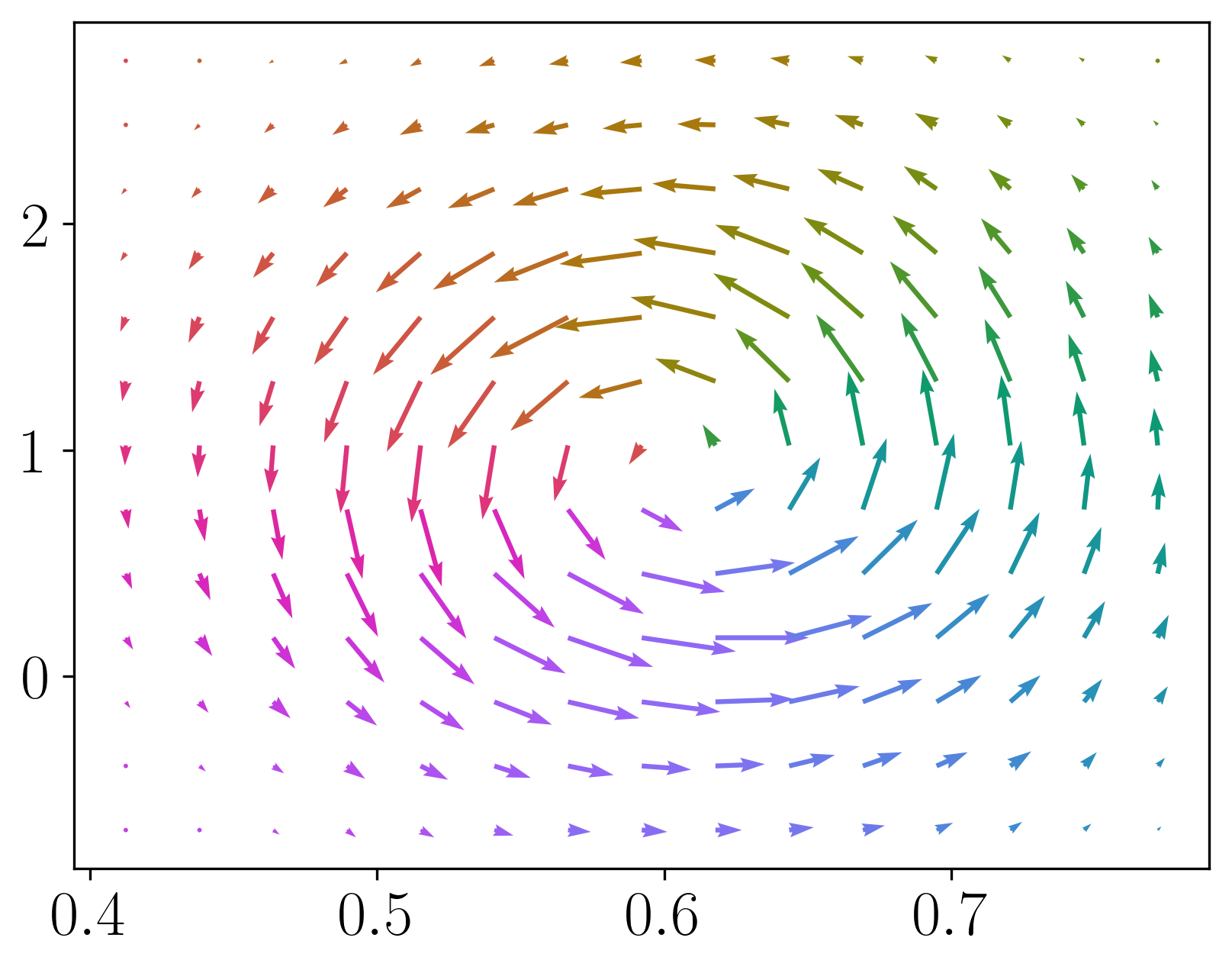}};
      \draw (2*\x-2,\yy) node{\includegraphics[width=\CurrentScale\textwidth]{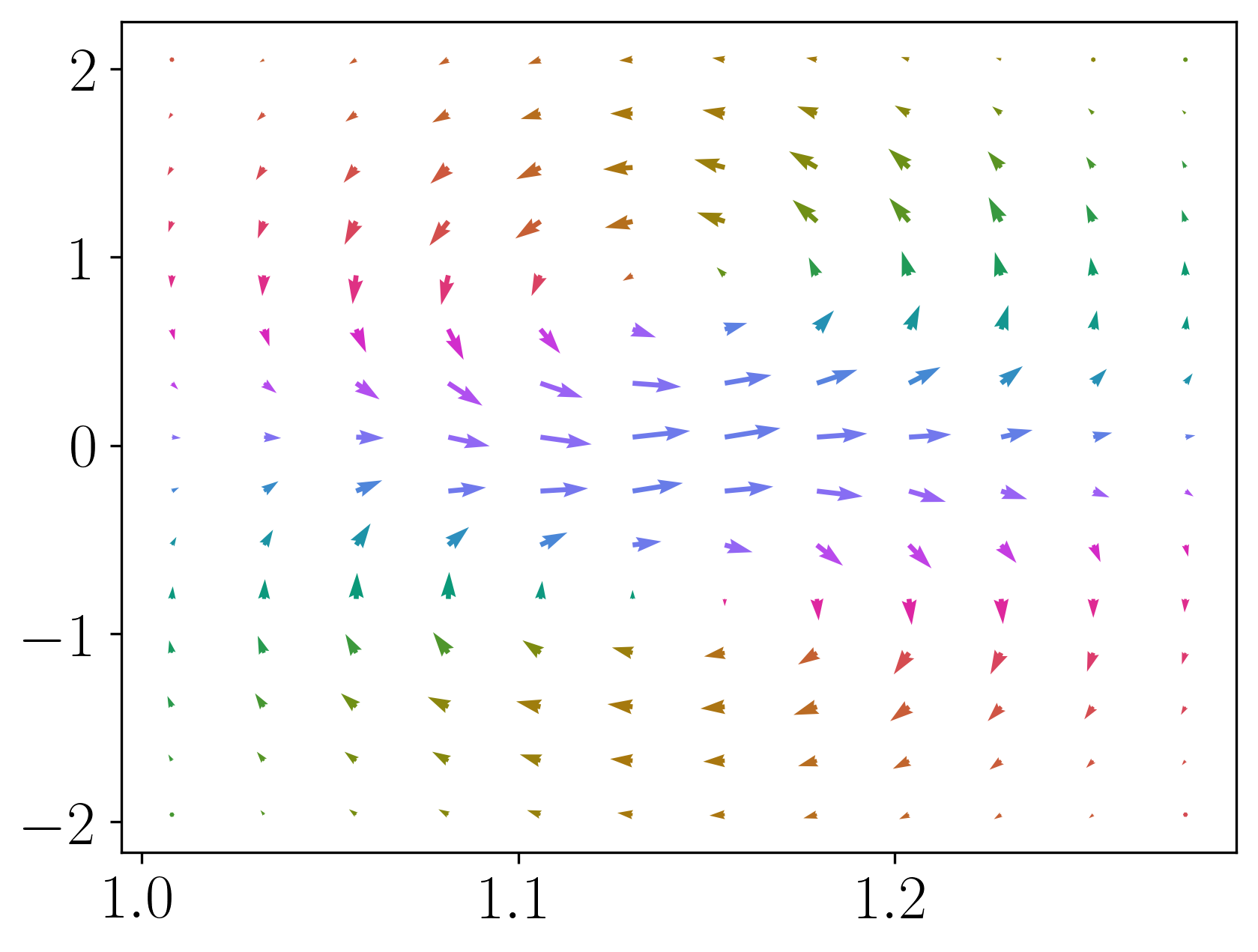}};
      \draw (3*\x-2+\xx,\yy) node{\includegraphics[width=\CurrentScale\textwidth]{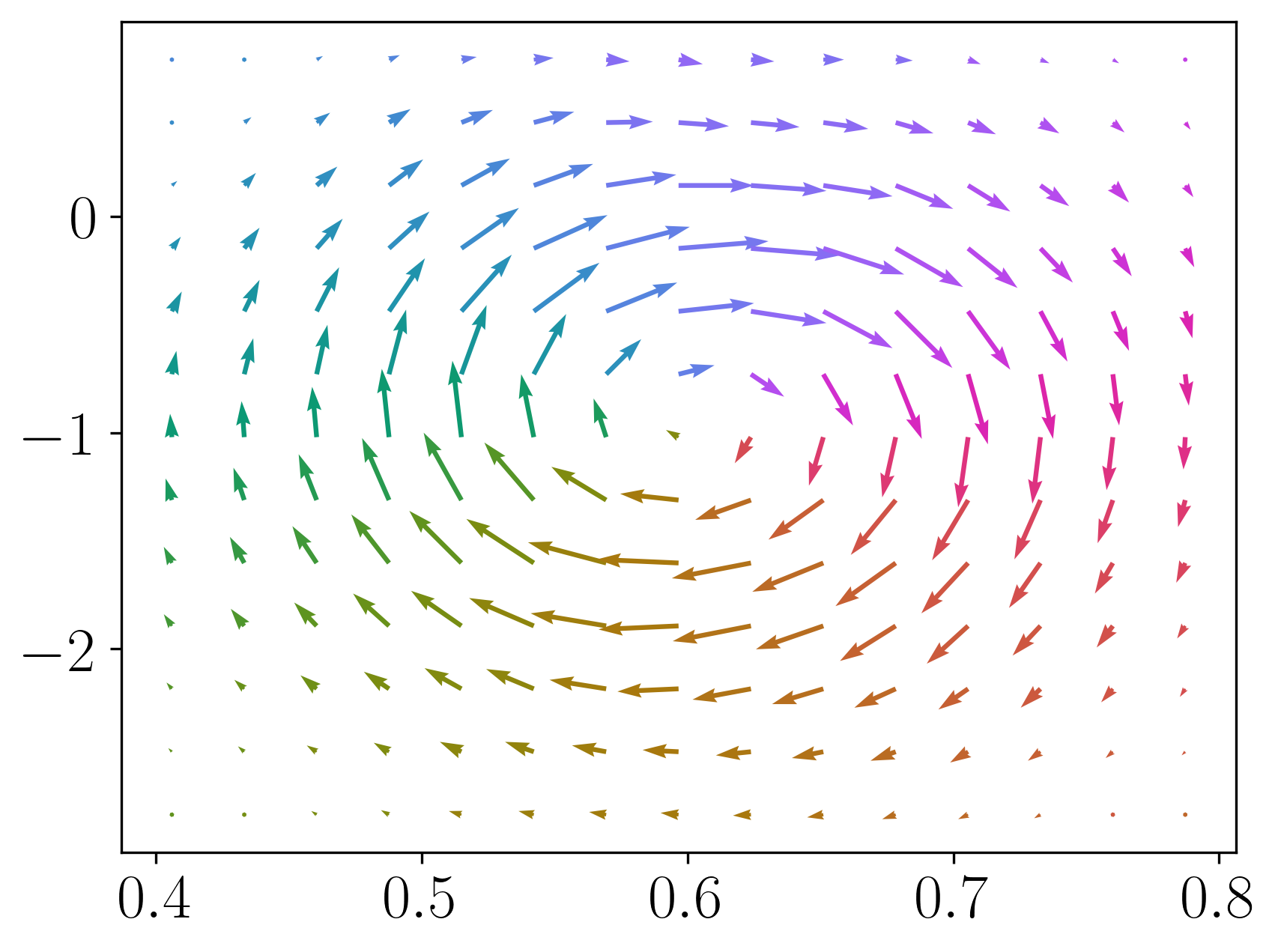}};
\draw (3*4.45 -0.8,\y) node{\includegraphics[width=.04\textwidth]{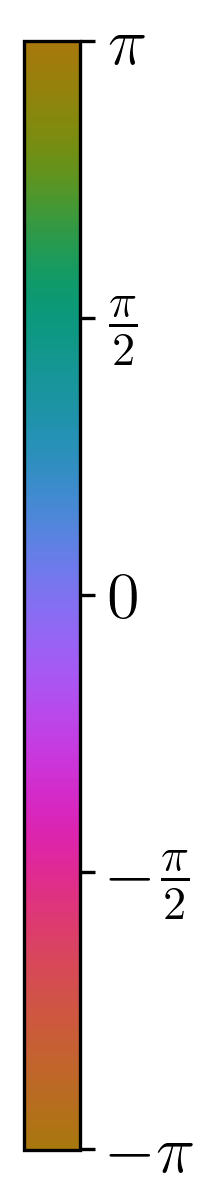}};

     \draw (\x-2-2.05,0) node[rotate=90] {$\partial_x\rho(\br_0)$};
     \draw (\x-2-2.05,\y) node[rotate=90] {$\partial_x\rho(\br_0)$};
     \draw (\x-2-2.05,\yy) node[rotate=90] {$\partial_x\rho(\br_0)$};
     \draw (\x-2+0.2,\yy-1.6) node {$\rho(\br_0)$};

     \draw (2*\x-2+0.2,\yy-1.6) node {$\rho(\br_0)$};
     
     \draw (3*\x-2+0.2,\yy-1.6) node {$\rho(\br_0)$};

\draw (\x-2+0.1,2.0-0.15) node {$\br_0=A$};
\draw (2*\x-2+0.1,2.0-0.15) node {$\br_0=B$};
\draw (3*\x-2+0.1,2.0-0.15) node {$\br_0=C$};
%

\draw [color={rgb, 255:red, 0; green, 0; blue, 0 }  ,draw opacity=1 ][line width=2]   (-4.61,0.05) -- (-1.485,0.05) ;
\draw (-4.2,0.13) node [anchor=north west][inner sep=0.75pt][circle,fill,inner sep=2pt,color=black][label = above:{\large \textcolor{black}{A}}]   [align=left] {};
\draw (-3.1,0.13) node [anchor=north west][inner sep=0.75pt][circle,fill,inner sep=2pt,color=black][label = above:{\large \textcolor{black}{B}}]   [align=left] {};
\draw (-2,0.13) node [anchor=north west][inner sep=0.75pt][circle,fill,inner sep=2pt,color=black][label = above:{\large \textcolor{black}{C}}]   [align=left] {};
%

\draw [color={rgb, 255:red, 0; green, 0; blue, 0 }  ,draw opacity=1 ][line width=2]   (-4.61,\y+0.1) -- (-1.485,\y+0.1) ;
\draw (-3.9,.18+\y) node [anchor=north west][inner sep=0.75pt][circle,fill,inner sep=2pt,color=black][label = above:{\large \textcolor{black}{A}}]   [align=left] {};
\draw (-3.1,0.18+\y) node [anchor=north west][inner sep=0.75pt][circle,fill,inner sep=2pt,color=black][label = above:{\large \textcolor{black}{B}}]   [align=left] {};
\draw (-2.3	,0.18+\y) node [anchor=north west][inner sep=0.75pt][circle,fill,inner sep=2pt,color=black][label = above:{\large \textcolor{black}{C}}]   [align=left] {};
%

\draw [color={rgb, 255:red, 0; green, 0; blue, 0 }  ,draw opacity=1 ][line width=2]   (-4.61,\yy+0.1) -- (-1.485,\yy+0.1) ;
\draw (-3.6,0.18+\yy) node [anchor=north west][inner sep=0.75pt][circle,fill,inner sep=2pt,color=black][label = above:{\large \textcolor{black}{A}}]   [align=left] {};
\draw (-3.1,0.18+\yy) node [anchor=north west][inner sep=0.75pt][circle,fill,inner sep=2pt,color=black][label = above:{\large \textcolor{black}{B}}]   [align=left] {};
\draw (-2.6,0.18+\yy) node [anchor=north west][inner sep=0.75pt][circle,fill,inner sep=2pt,color=black][label = above:{\large \textcolor{black}{C}}]   [align=left] {};

\draw (-0.75,1.7) node [anchor=north west][inner sep=0.75pt] [align=left] {$\rho$};
\draw (-0.75,\y+1.7) node [anchor=north west][inner sep=0.75pt] [align=left] {$\rho$};
\draw (-0.75,\yy+1.6) node [anchor=north west][inner sep=0.75pt] [align=left] {$\rho$};

\end{tikzpicture}
\end{center}
\caption{\label{FigCurrents} Measurements of the stationary
  probability currents in the planes
  $(\rho(\br_0),\partial_x\rho(\br_0))$ at representative points in
  phase-separated systems, using numerical resolution of
  Eq.~\eqref{SPDE}. The rows corresponds to $2\lambda+\kappa'<0$
  (top), $2\lambda+\kappa'=0$ (center), and $2\lambda+\kappa'>0$
  (bottom), respectively. The average stationary profiles are shown in
  the left column. Other columns show the current vector fields
  measured at the corresponding points $\br_0= A, B, C$ in the
  phase-separated profiles. (Arrow colors encode their angles with
  respect to $\be_x$.) Parameters: $a=-1$, $b=1$, $\kappa=0.1$,
  average density $\rho_0=-0.4$, $D=10^{-3}$, $L_x=L_y=10$ and
  $\lambda=-2$ (top), 0 (center) and 2 (bottom). See~\cite{supp} for
  numerical details.}
\end{figure*}

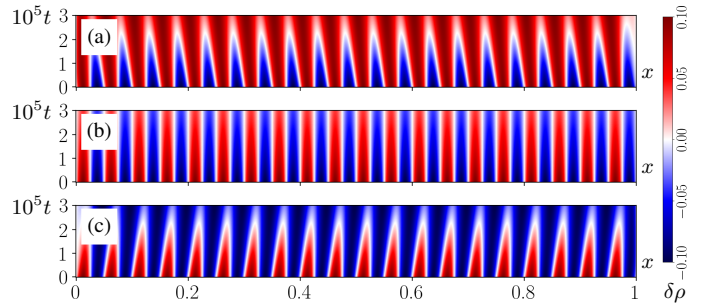
\begin{figure}[b!]
\begin{center}
\hbox{\hspace{-1.5em}
\begin{tikzpicture}[thick,scale=0.9, every node/.style={scale=0.9}]	
	  \def\w{0.49}
      \def\Ox{0}
      \def\x{6.2}
      \def\y{-1.4}
      \def\yy{-0.75}
      \def\xx{-3.2}

\draw (\Ox+4.8,-1.2) node[rotate=90] {\includegraphics[width=0.23\textwidth]{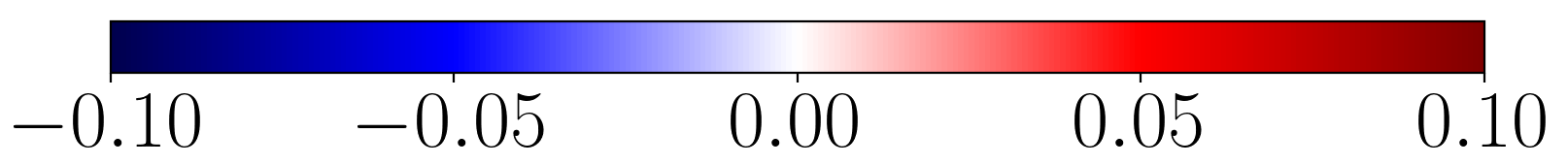}}; 
\draw (4.8,-3.8) node[label = above:{$\delta\rho$}]   [align=left] {};
%
\draw (\Ox,0) node {\includegraphics[width=\w\textwidth]{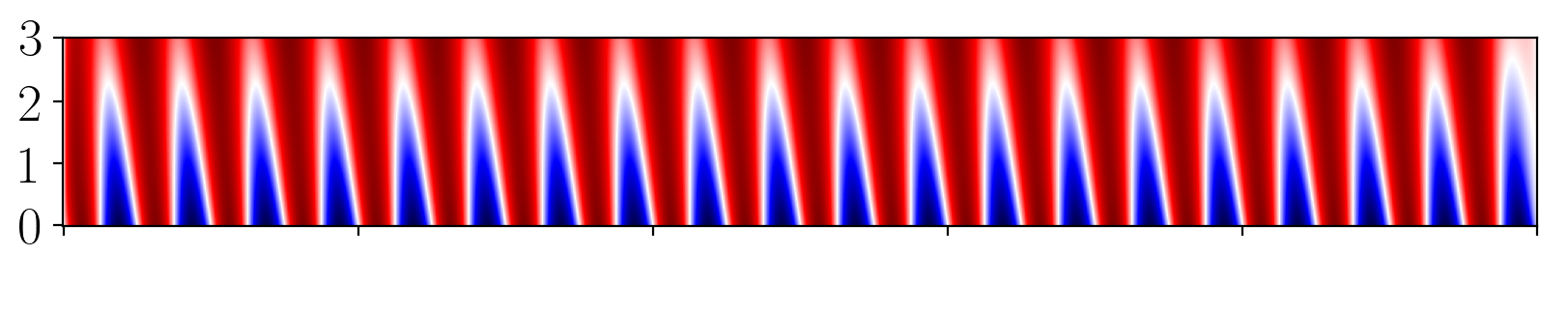}};
\draw (\Ox,\y) node{\includegraphics[width=\w\textwidth]{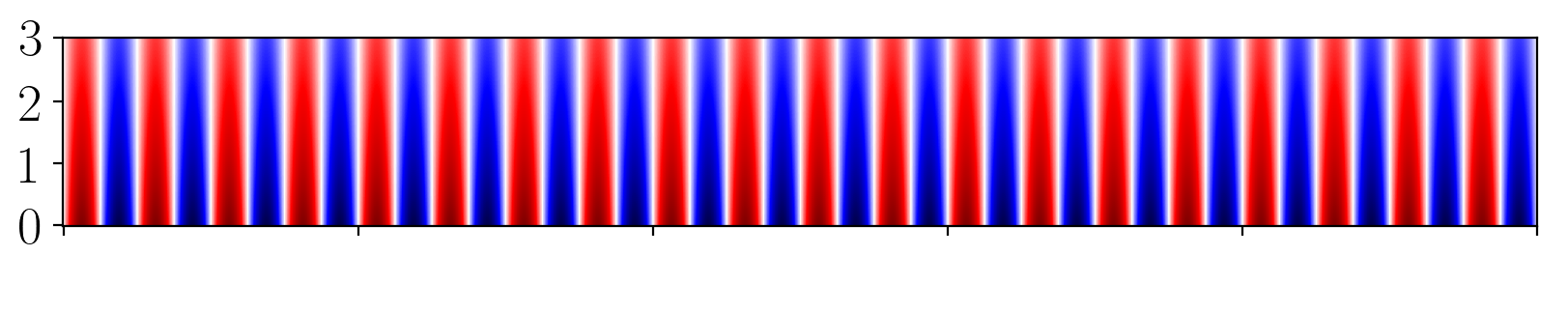}};
\draw (\Ox,2*\y) node{\includegraphics[width=\w\textwidth]{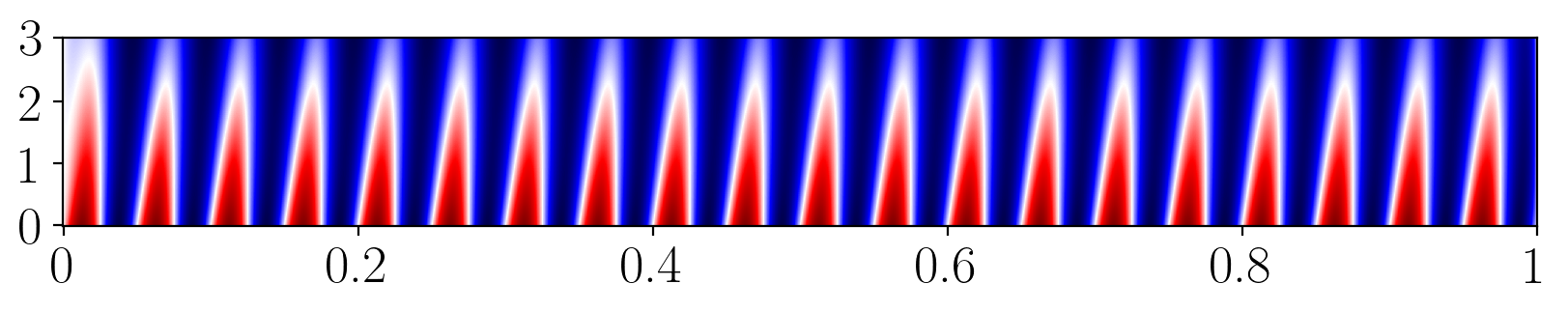}};
\begin{scope}[xshift=-0.5cm,yshift=0.1cm]
\filldraw[white] (-0.25+\xx,.75+\yy) rectangle (0.25+\xx,1.25+\yy);
\draw (\xx,1+\yy) node {(a)};
\end{scope}

\begin{scope}[xshift=-0.5cm,yshift=0.1cm]
\filldraw[white] (-0.25+\xx,.75+\yy+\y) rectangle (0.25+\xx,1.25+\yy+\y);
\draw (\xx,1+\yy+\y) node {(b)};
\end{scope}

\begin{scope}[xshift=-0.5cm,yshift=0.1cm]
\filldraw[white] (-0.25+\xx,.75+\yy+2*\y) rectangle (0.25+\xx,1.25+\yy+2*\y);
\draw (\xx,1+\yy+2*\y) node {(c)};
\end{scope}

\draw (-4.7,0.3) node[label = above:{$10^5t$}]   [align=left] {};
\draw (-4.7,0.3+\y) node[label = above:{$10^5t$}]   [align=left] {};  
\draw (-4.7,0.3+2*\y) node[label = above:{$10^5t$}]   [align=left] {};

\draw (4.4,-0.5) node[label = above:{$x$}]   [align=left] {};
\draw (4.4,-0.5+\y) node[label = above:{$x$}]   [align=left] {};  
\draw (4.4,-0.5+2*\y) node[label = above:{$x$}]   [align=left] {};
\end{tikzpicture}
}
\end{center}
\vspace*{-3em}
\caption{Kymographs showing the zero-noise relaxation of a
  perturbation $\delta h(x)=0.1 \sin(40\pi x)$, added at time $t=0$
  to a linear profile $h(x)=10 x$, under the KPZ
  dynamics~\eqref{eq:KPZ} with Dirichlet boundary conditions. Parameters: $\nu=2$ and $\lambda=-16, 0,
  16$ for panels (a), (b) and (c), respectively. System size
  $L=1$. Space and time discretization: $\rmd x=10^{-3}$ and $\rmd
  t=10^{-7}$.}\label{PropagWavesKPZ}
\end{figure}

In this Letter we address this challenge by introducing a new
mathematical framework that enables a systematic characterization of
steady-state probability currents in nonequilibrium stochastic field
theories. This framework is based on a generalization of the curl
operator to functional spaces in the form of a functional exterior
derivative and on the identification of the appropriate Riemannian
metric on the space of fields. We note that a related object, called
`vertical derivative', has been introduced for jet
bundles~\cite{anderson1989variational}, a context more restrictive
than what we present here. In addition, differential geometry has been
formally extended to abstract mathematical
spaces~\cite{kriegl1997convenient} but the corresponding level of
abstraction makes such theory hardly applicable to concrete physics
problems~\cite{supp}. Furthermore, these mathematical formalisms have
never been applied to characterize probability currents in stochastic
field theories.
Below, we briefly recap the finite-dimensional case to highlight the
key steps of its generalization to infinite dimension. We then detail
the construction of the functional exterior derivative for two
important examples: the Active Model B
(AMB)~\cite{wittkowski2014scalar} and the Kardar-Parisi-Zhang (KPZ)
equation~\cite{kardar1986dynamic}.  Importantly, when undergoing motility-induced phase separation (MIPS),
AMB leads to a finite entropy production rate localized at the
liquid-gas interface~\cite{nardini2017entropy,
  martin2008statistical}. However, the corresponding probability
currents have remained out of reach so far.
Here, we show how these currents can be decomposed into superpositions
of local 2D rotations, allowing for direct observation (see
Fig.~\ref{FigCurrents}). Our framework also reveals the direct
manifestations of these high-dimensional currents in real space, in
the form of propagating modes localized at the liquid-gas interface
(see Fig.~\ref{PropagWaves}). Similarly, for the KPZ equation, we show
how fluctuations are advected along height gradients (see
Fig.~\ref{PropagWavesKPZ}).

To set the stage for stochastic field theories, we start with a quick
reminder of the well-known finite-dimensional case.  Consider the $n$-dimensional Langevin dynamics
\begin{equation}
\dot{\br}(t)=\bF(\br(t)) + \sqrt{2D} \boldeta(t) \: ,
\label{EDS}
\end{equation}
where $\br(t)\in\mathbb{R}^n$, $\boldeta$ is a Gaussian white noise of
zero mean and unit variance, $D$ is the diffusion constant, the
mobility has been set to 1, and $\bF$ is an arbitrary smooth vector
field. The corresponding Fokker-Planck equation reads
$\partial_t p = -\nabla\cdot\bJ$, with $\bJ=p \bF -D\nabla p$.
In the steady state, the probability current $\bJ_s$ encodes the
advection of the probability $p_s$ by the velocity field
$\bv_s \equiv \bJ_s/p_s = \bF - D\nabla\log p_s $. The flow lines of
$\bv_s$ indicate the typical trajectories of the system in the steady
state~\cite{liverpool2018non}. Because it favors certain trajectories
over their time-reversed counterparts, the swirling behavior of
$\bv_s$ is responsible for the irreversibility of
dynamics~\eqref{EDS}.  When $n=3$, it is characterized by the
vorticity
$\bomega(\br)\equiv \nabla\times \bv_s(\br)=\nabla\times \bF(\br)$
whose norm gives the angular speed of the local swirls, and whose direction is orthogonal to the local 2-dimensional planes in which the current undergoes local rotations.
Further, note that the entropy production rate of dynamics~\eqref{EDS}
is given by $\sigma = D^{-1}\int \bJ_s\cdot\bF \: \rmd\br$
\cite{seifert2005entropy}. Since $\bJ_s$ is divergence free and
$\mathbb{R}^n$ simply connected, there is a vector field $\bC$
such that $\bJ_s=\nabla\times\bC$. Integrating by parts,
one gets $\sigma=D^{-1}\int \textbf{C}(\br)\cdot\bomega(\br) \: \rmd\br $.
Hence, $\bomega(\br)$ can be seen as the local source of entropy
production and $\bC(\br)$ as a weight over the infinitesimal loops
around $\br$.

The generalization to arbitrary finite dimension $n$ amounts to
replacing $\nabla\times \bF$ by $\rmd \bF^\flat$ in
the vorticity $\bomega$, where $\rmd$ is the exterior derivative and
$\bF^\flat$ the one-form associated to $\bF$ through a Riemannian
metric $g$~\cite{jiang2004mathematical}. Denote by
$(\be_i)_{i=1,...,d}$ a local basis and $(\rmd x^i)_{i=1,...,d}$ its
dual, which satisfies $\rmd x^i(\be_j)=\delta^i_j$. Then, to
 $\bF=\sum_i F^i\be_i$, we associate the one-form
$\bF^\flat = \sum_i F_i\rmd x^i = \sum_{i,j} g_{ij}F^j\rmd x^i$ with
$g_{ij}\equiv g(\be_i,\be_j)$. The exterior derivative of $\bF^\flat$
is then the two-form $\rmd\bF^\flat$ whose action on arbitrary pairs
$\bu,\bv$ of vector fields reads
\begin{equation}
\mathrm{d}\bF^\flat(\bu,\bv) = \sum_{i,j=1}^n \Big(\frac{\partial F_j}{\partial x_i} - \frac{\partial F_i}{\partial x_j}\Big)u^iv^j \ .
\label{ExtDivExplicite0}
\end{equation}
Denoting by $\mathrm{d}x^i\wedge\mathrm{d}x^j$ the bilinear maps such that $\mathrm{d}x^i\wedge\mathrm{d}x^j(\bu,\bv) = u^iv^j-u^jv^i$, the vorticity reads
\begin{equation}
\bomega\equiv \mathrm{d}\bF^\flat = \sum_{1\leq i<j\leq n} \Big(\frac{\partial F_j}{\partial x_i} - \frac{\partial F_i}{\partial x_j}\Big)\mathrm{d}x^i\wedge\mathrm{d}x^j \ .
\label{ExtDivExplicite}
\end{equation}
The prefactor of $\rmd x^i\wedge \rmd x^j$ in
Eq.~\eqref{ExtDivExplicite} measures the local rotation induced by
$\bF$ in the $(\be_i,\be_j)$ plane: its sign gives the direction of
the rotation and its amplitude the angular speed.
Finally, note that dynamics~\eqref{EDS} is reversible if and only if
$\rmd\bF^\flat=0$. Then, $\bF$ is a gradient and Eq.~\eqref{EDS} is a
stochastic gradient descent.

We now turn to the core results of this Letter: the generalization of
Eq.~\eqref{ExtDivExplicite} to infinite-dimensional stochastic field
theory and the physical insight it provides on the corresponding
systems.
For an arbitrary field theory, this requires generalizing the $\flat$
and $\rmd$ operators. The former amounts to finding the Riemannian
metric that identifies a reversible dynamics with a gradient descent of the free-energy;
it also associates a one form to the deterministic drift. The exterior
derivative then extracts the skew-symmetric part of the corresponding
Jacobian, which vanishes for equilibrium dynamics and identifies the
non-equilibrium circulations otherwise. In this Letter, for sake of
clarity, we present these ideas on two important examples, AMB and the
KPZ equation. Details of the underlying
construction are provided in~\cite{supp}.

\textit{Active Model B.} We start with the study of AMB, which has
attracted a lot of interest
recently~\cite{nardini2017entropy,wittkowski2014scalar}, and whose
probability currents have remained elusive so far. AMB describes a
scalar field whose dynamics is given by
\begin{equation}
\partial_t\rho = -\nabla\cdot \bj \: , \: \text{where} \:\: \bj = - \nabla \mu + \sqrt{2D} \bLambda\;,
\label{SPDE}
\end{equation} 
where $\bLambda(\br,t)$ is a centered Gaussian white noise field of unit variance and $\mu$ a nonequilibrium chemical potential defined by 
\begin{equation}
  \mu([\rho],\br) = a\rho + b\rho^3 -\kappa(\rho)\Delta\rho +\lambda(\rho) \vert\nabla\rho\vert^2 \: .
\label{ChemicalPotential}
\end{equation}
Note that $a$, $b$ and $D$ are constants but $\lambda$ and
$\kappa$ depend on $\rho(\br)$.

To carry out the aforementioned geometric program, as detailed
in~\cite{supp}, we first introduce the space of vector
fields~$\bmu(\br,[\rho])$ generated by all chemical potentials
$\mu(\br,[\rho])$ through
\begin{equation}
\bmu(\br,[\rho]) \equiv -\Delta \mu(\br,[\rho]) \ .
\end{equation} 
This allows rewriting Eq.~\eqref{SPDE} as
$\partial_t\rho = -\bmu+\nabla \cdot \sqrt{2D}\bLambda$ and we now
turn to construct the vorticity of $-\bmu$. Following
Otto~\cite{otto2001geometry}, we define the Riemannian metric
\begin{equation}
g_\rho(\bmu_1,\bmu_2) \equiv \int \nabla \mu_1(\br,[\rho])\cdot \nabla \mu_2(\br,[\rho]) \rmd \br \ .
\label{eq:RiemMetric}
\end{equation}
As in finite dimension, this  allows associating to any vector
field $\bmu_1$ a funtional one-form, $\bmu_1^\flat$, through
$\bmu_1^\flat(\bmu_2) \equiv g(\bmu_1,\bmu_2)$. An integration by
parts in Eq.~\eqref{eq:RiemMetric} gives
\begin{equation}
\bmu_1^\flat(\bmu_2)  = \int \mu_1(\br,[\rho]) \bmu_2(\br,[\rho])\rmd \br \ . 
\end{equation}
By analogy with the finite-dimensional case~\eqref{ExtDivExplicite0}, we define the functional exterior derivative of any one-form $\bmu^\flat$ through its action on an arbitrary pair $\bphi,\bpsi$ of vector fields:
\begin{equation}
\mbd \bmu^\flat (\bphi,\bpsi) = \int \left[\frac{\delta\mu(\br,[\rho])}{\delta\rho(\br')}-\frac{\delta\mu(\br',[\rho])}{\delta\rho(\br)}\right] \bphi(\br')\bpsi(\br) \rmd \br\rmd\br' \ .
\label{eq:FunctExtDer}
\end{equation}
For the chemical potential~\eqref{ChemicalPotential}, integrating by parts and rearranging the terms leads to:
\begin{equation}
\mbd \bmu^\flat (\bphi,\bpsi) = \int (2\lambda+\kappa')\nabla\rho\cdot(\bpsi\nabla\bphi-\bphi\nabla\bpsi) d\br
\end{equation}
where $\kappa'(\rho)\equiv\frac{d \kappa(\rho)}{d\rho}$.  Finally, $\mbd\bmu^\flat$ can be rewritten as
\begin{equation}
\mathbb{d} \bmu^\flat = \int \rmd\br (2\lambda + \kappa')\nabla \rho\cdot \delta_\br \wedge \nabla\delta_\br \: ,
\label{dmu}
\end{equation}
where $\delta_\br$ is the Dirac delta at position $\br$ and $\wedge$
the extension of the finite-dimensional wedge product to distributions
such that
$\delta_\br\wedge\nabla\delta_\br (\bphi,\bpsi) \equiv
\bpsi(\br)\nabla\bphi(\br)-\bphi(\br)\nabla\bpsi(\br)$.

The vorticity $\bomega$ of the deterministic drift of Eq.~\eqref{SPDE}
is then given by $\bomega \equiv -\mbd \bmu^\flat$.
Equation~\eqref{eq:FunctExtDer} shows that $\bomega=0$ corresponds to
the Schwarz condition for $\mu$ to be the functional derivative of a
free energy~\cite{o2020lamellar,grafke2017spatiotemporal}. For AMB,
this amounts to
$2\lambda
+\kappa'=0$~\cite{wittkowski2014scalar,solon2018generalized}.%

Comparing Eq.~\eqref{dmu} to Eq.~\eqref{ExtDivExplicite}, the discrete
sum over $\mathrm{d}x_i\wedge\mathrm{d}x_j$ has been replaced by an
integral  over $\delta_\br\wedge\nabla\delta_\br$.
Equation \eqref{dmu} can thus be interpreted as follows: the flow
lines of $-\bmu$ swirl around a given point $\rho$ in $\mbF$
as soon as $(2\lambda+\kappa')\nabla\rho \neq 0$.  As in the
finite-dimensional setting, such a local swirl corresponds to an
infinitesimal rotation that can be decomposed into the superposition
of rotations occurring in the spaces
$(\rho(\br),\partial_{x}\rho(\br),\partial_{y}\rho(\br))$ wherever
$(2\lambda+\kappa')\nabla\rho(\br) \neq 0$.
All in all,
the flow lines of the deterministic drift tend to rotate in the 2D plane
$(\rho(\br),\partial_k\rho(\br))$:
\begin{eqnarray}
\left\{
    \begin{array}{ll}
        \text{counter-clockwise} & \mbox{iff } \quad [2\lambda+\kappa']\partial_{k}\rho(\br)>0 \\
        \text{clockwise} & \mbox{iff } \quad [2\lambda+\kappa']\partial_{k}\rho(\br) < 0
    \end{array}
\right.
\label{CurrentSwirls}
\end{eqnarray}
at speed given by the amplitude of $(2\lambda+\kappa')\partial_{k}\rho(\br)$.

Let us now show that these predictions allow measuring the
steady-state currents of a phase-separated AMB. We denote by
$\rho_s(\br)\equiv \langle\rho(\br)\rangle$ the stationary average
profile of the fluctuating field $\rho$. We then measure the
probability current in the plane $(\rho(\br),\partial_x\rho(\br))$ at
three different positions (points A, B and C - see
Fig.~\ref{FigCurrents}) along the horizontal diameter of the liquid
droplet.
As predicted, changing the sign of $2\lambda+\kappa'$ (top vs. bottom row of Fig.~\ref{FigCurrents}) or that of $\partial_x\rho$ (column A vs. C) changes the direction of the circulation. Furthermore, the probability current vanishes in the equilibrium case $2\lambda + \kappa'=0$ (center row), as expected. 

Note that, at the interface, $\rho(\br)\simeq\rho_s(\br)$ and $\nabla\rho(\br)\simeq\nabla\rho_s(\br)$ so that $\left.\mbd\bmu^\flat\right|_\rho\simeq\left.\mbd\bmu^\flat\right|_{\rho_s}$, which corresponds to uniform rotations in each space $(\rho(\br),\nabla\rho(\br))$. The latter give rise to the leading order terms of the current in the noise amplitude (columns A and C). On the contrary, in the bulk,	$\nabla\rho(\br)\simeq\nabla\delta\rho(\br)$, where $\delta\rho\equiv\rho-\rho_s$. This leads to weaker, higher order currents (column B).

It is tempting to split the chemical potential into $\mu=\mu_{\rm eq} + \mu_{\rm act}$, where $\mu_{\rm eq}$ is the functional derivative of a free energy $\mcF$ whereas $\mu_{\rm act}$ is not integrable, so as to identify $\mu_{\text{act}}$ as the source of irreversibility. Unfortunately, such a decomposition is not unique, since adding a functional derivative $\delta\mathcal{G}/\delta\rho$ to $\mu_{\text{eq}}$ and subtracting it from $\mu_{\text{act}}$ yields another equivalent decomposition. On the contrary, $\mathbb{d}\bmu^\flat$ can unambiguously be identified as the source of irreversibility since the set of functional derivatives exactly coincides with the kernel of $\mathbb{d}$, so that 
\begin{equation}
\mathbb{d}\bmu^\flat=\mathbb{d}\bmu^\flat_{\rm act}=\mathbb{d}(\bmu^\flat_{\rm act} + \delta\mathcal{G}/\delta\rho) \ .
\label{eq:unambiguousTRSBmeasure}
\end{equation}
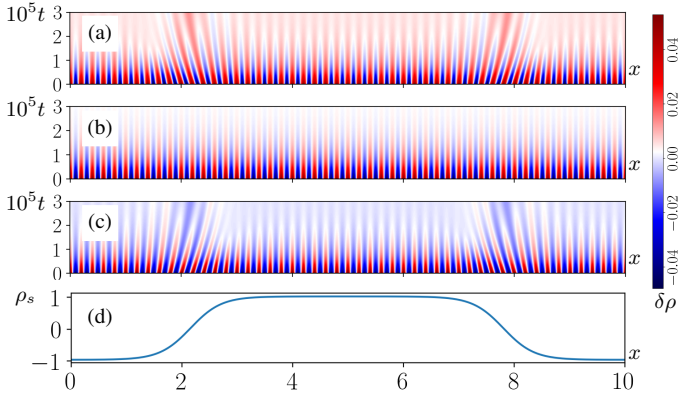
\begin{figure}
\begin{center}
\hbox{\hspace{-1.5em}
\begin{tikzpicture}	[thick,scale=0.9, every node/.style={scale=0.9}]	
	  \def\w{0.49}
      \def\Ox{0}
      \def\x{6.2}
      \def\y{-1.4}
      \def\yy{-0.75}
      \def\xx{-3.1}

\draw (\Ox+4.75,-1.4) node[rotate=90] {\includegraphics[width=0.23\textwidth]{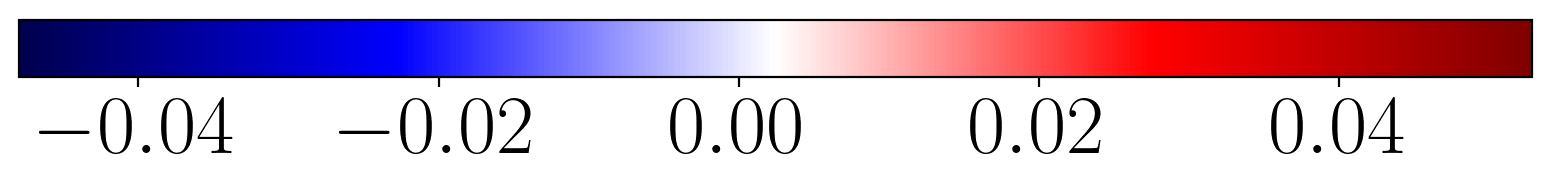}}; 
\draw (\Ox+4.75,-4) node[label = above:{$\delta\rho$}]   [align=left] {};

\draw (\Ox,0) node {\includegraphics[width=\w\textwidth]{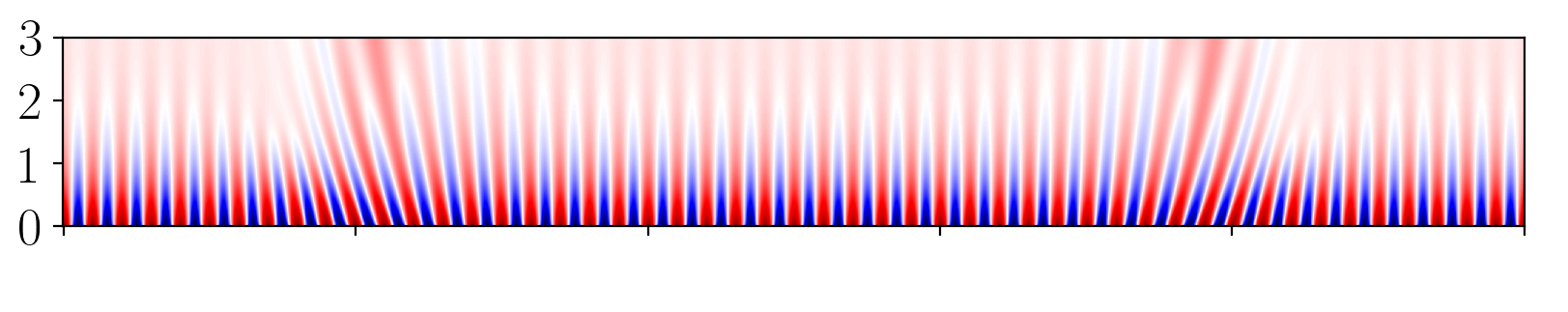}};
\draw (\Ox,\y) node{\includegraphics[width=\w\textwidth]{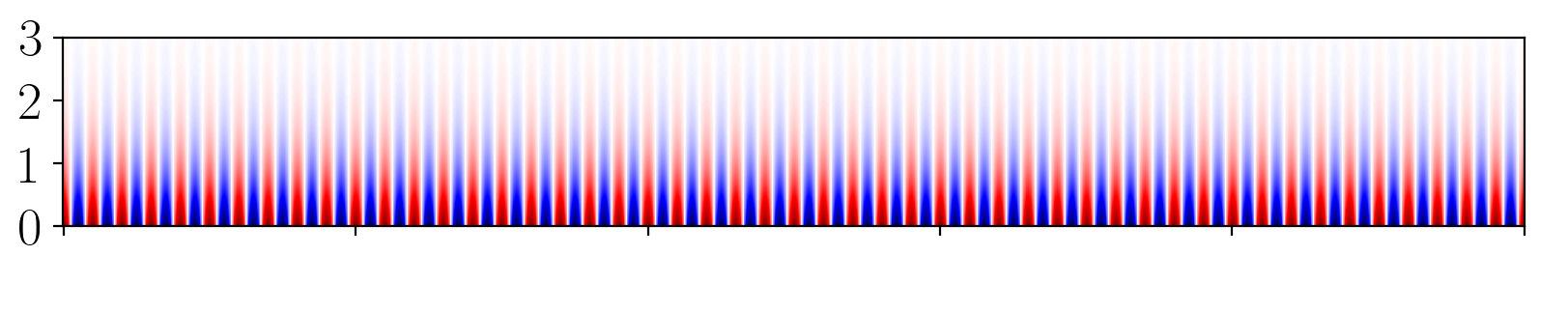}};
\draw (\Ox,2*\y) node{\includegraphics[width=\w\textwidth]{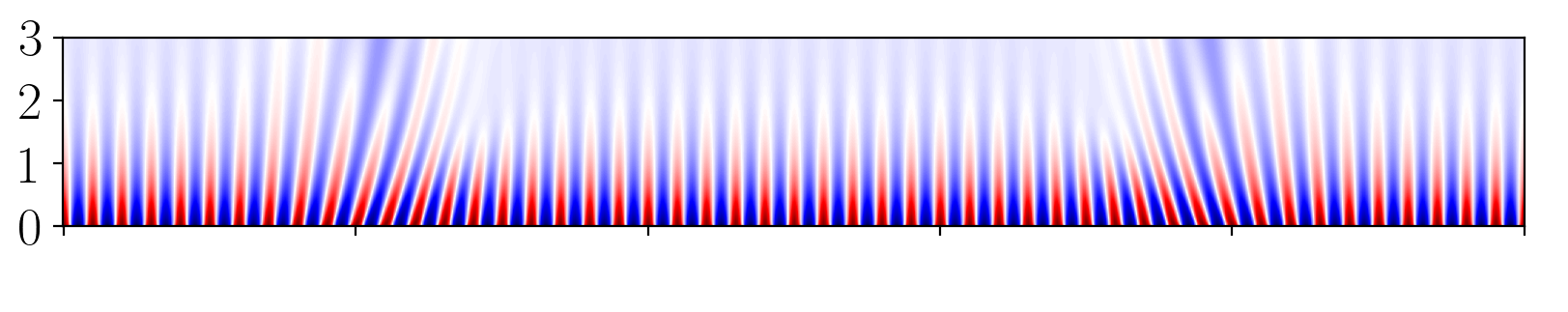}};
\draw (\Ox-0.15,3*\y) node{\includegraphics[width=0.512\textwidth]{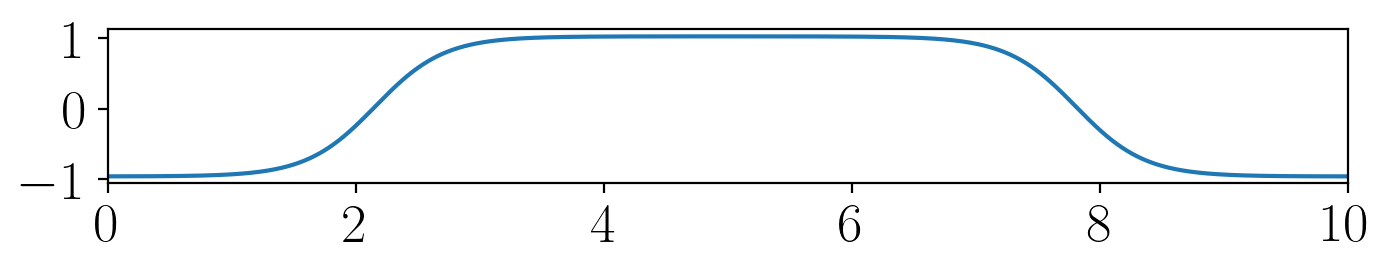}};
\begin{scope}[xshift=-0.5cm,yshift=0.1cm]
\filldraw[white] (-0.25+\xx,.75+\yy) rectangle (0.25+\xx,1.25+\yy);
\draw (\xx,1+\yy) node {(a)};
\end{scope}

\begin{scope}[xshift=-0.5cm,yshift=0.1cm]
\filldraw[white] (-0.25+\xx,.75+\yy+\y) rectangle (0.25+\xx,1.25+\yy+\y);
\draw (\xx,1+\yy+\y) node {(b)};
\end{scope}

\begin{scope}[xshift=-0.5cm,yshift=0.1cm]
\filldraw[white] (-0.25+\xx,.75+\yy+2*\y) rectangle (0.25+\xx,1.25+\yy+2*\y);
\draw (\xx,1+\yy+2*\y) node {(c)};
\end{scope}

\begin{scope}[xshift=-0.5cm,yshift=0.1cm]
\filldraw[white] (-0.25+\xx,.75+\yy+3*\y) rectangle (0.25+\xx,1.25+\yy+3*\y);
\draw (\xx,1+\yy+3*\y) node {(d)};
\end{scope}

\draw (-4.7,0.3) node[label = above:{$10^5t$}]   [align=left] {};
\draw (-4.7,0.3+\y) node[label = above:{$10^5t$}]   [align=left] {};  
\draw (-4.7,0.3+2*\y) node[label = above:{$10^5t$}]   [align=left] {};
\draw (-4.7,0.3+3*\y) node[label = above:{$\rho_s$}]   [align=left] {};

\draw (4.3,-0.5) node[label = above:{$x$}]   [align=left] {};
\draw (4.3,-0.5+\y) node[label = above:{$x$}]   [align=left] {};  
\draw (4.3,-0.5+2*\y) node[label = above:{$x$}]   [align=left] {};
\draw (4.3,-0.5+3*\y) node[label = above:{$x$}]   [align=left] {};

\end{tikzpicture}
}
\end{center}
\vspace*{-2em}
\caption{Evolution of a perturbation $\delta\rho$ around the
  equilibrium profile $\rho_s$ under the AMB dynamics~\eqref{SPDE}
  using periodic boundary conditions and
  $\delta\rho(x,y,t=0)=\varepsilon\cos(100\pi x/L_x)$. Panels (a-c) are
  kymographs representing the evolution of $\delta\rho(x,L_y/2,t)$. Panel (d)
  is a cut of $\rho_s$ at $y=L_y/2$. Parameters: $\rho_0=-0.45$,
  $a=-1$, $b=1$, $\kappa=0.15$, $dt=10^{-7}$, $dx=dy=10^{-2}$,
  $L_x=L_y=10$, $\varepsilon=0.05$, and $\lambda=-4$, 0 and 4 for
  panels (a), (b) and (c), respectively. Time axis unit of the kymographs is $\Delta t =10^{-5}$.}\label{PropagWaves}
\end{figure}

\begin{figure}
\begin{center}
\begin{tikzpicture}[thick,scale=0.8, every node/.style={scale=0.8}]
      \def\w{0.45}

      \def\xa{4.4}
      \def\ya{-0.4}
      \def\xb{1.2}
      \def\yb{-3.8}
      \def\xc{-4.35}
      \def\yc{-0.4}
      \def\xd{-1.2}
      \def\yd{3.05}
      
      \def\xxa{3.3}
      \def\yya{-1.15}
      \def\xxb{0.25}
      \def\yyb{-2.9}
      \def\xxc{-3.3}
      \def\yyc{0.45}
      \def\xxd{-0.25}
      \def\yyd{2.8}

\draw (0,0) node {\includegraphics[width=\w\textwidth]{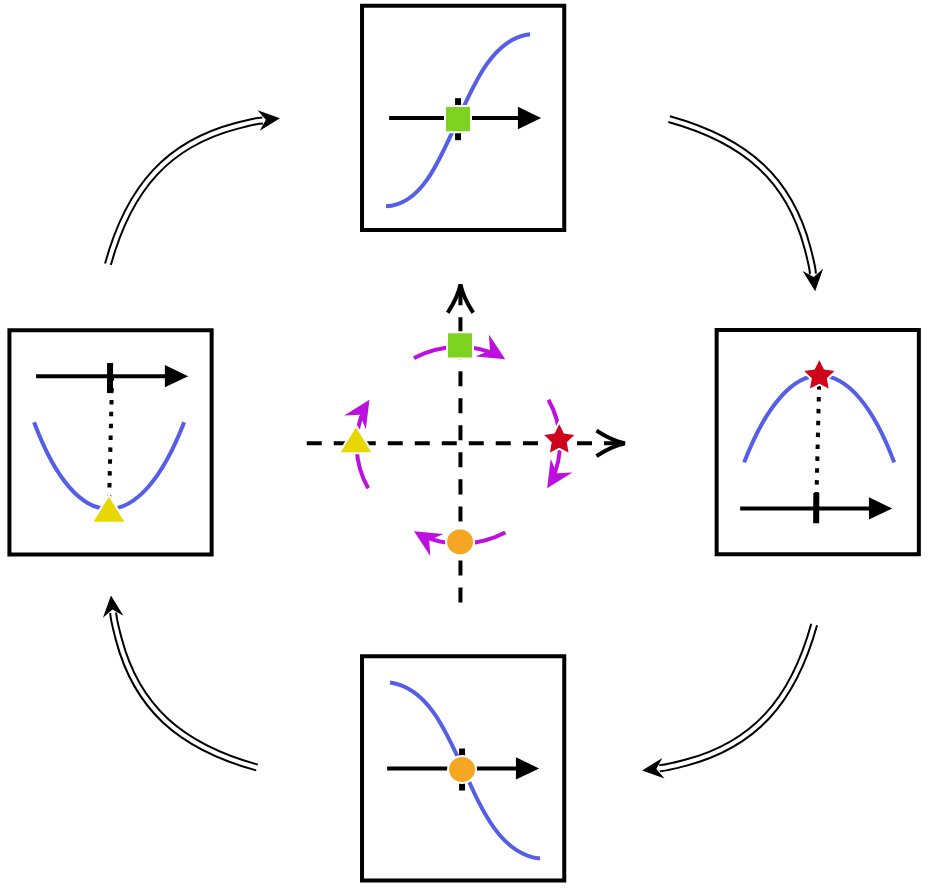}};

\draw (\xa,\ya) node[label = above:{(1)}]   [align=left] {};
\draw (\xb,\yb) node[label = above:{(2)}]   [align=left] {};  
\draw (\xc,\yc) node[label = above:{(3)}]   [align=left] {};
\draw (\xd,\yd) node[label = above:{(4)}]   [align=left] {};

\draw (0.1,1.5) node [anchor=north west][inner sep=0.75pt] [align=left] {\large $\partial_x\delta\rho(\br_0)$};
\draw (1,-0.15) node [anchor=north west][inner sep=0.75pt] [align=left] {\large $\delta\rho(\br_0)$};

\draw (\xxa-0.4,\yya) node[label = above:{\large $x_0$}]   [align=left] {};
\draw (\xxb,\yyb) node[label = above:{\large $x_0$}]   [align=left] {};  
\draw (\xxc,\yyc) node[label = above:{\large $x_0$}]   [align=left] {};
\draw (\xxd,\yyd) node[label = above:{\large $x_0$}]   [align=left] {};

\draw (\xxa+0.5,\yya+0.1) node[label = above:{\large $x$}]   [align=left] {};
\draw (\xxb+0.45,\yyb-0.4) node[label = above:{\large $x$}]   [align=left] {};  
\draw (\xxc+0.9,\yyc-0.4) node[label = above:{\large $x$}]   [align=left] {};
\draw (\xxd+0.95,\yyd-0.45) node[label = above:{\large $x$}]   [align=left] {};

\end{tikzpicture}
\end{center}
\caption{Analysis of the currents shown in Fig.~\ref{FigCurrents}
  predicting the mode propagation reported in
  Fig.~\ref{PropagWaves}. Consider the case $2\lambda + \kappa'<0$ and
  a point $\br_0=(x_0,L_y/2)$ on the left boundary of the droplet (top
  row, column A of Fig.~\ref{FigCurrents}). In the central panel
  above, we show the circulation induced by the current in the
  $(\delta\rho(\br_0),\partial_x\delta\rho(\br_0))$ plane. Consider a
  perturbation such that $\delta\rho(\br_0)$ is a local maximum at
  $t=0$ (panel (1), red star). As time goes on, the current drives the
  fluctuation sequentially from (1) to (2) (orange dot), to (3)
  (yellow triangle), to (4) (green square). Each panel shows the 
  fluctuation profile $\delta\rho(x,L_y/2)$ around $x_0$ (blue
  curves). These successive states of $\delta\rho$ shows that
  the probability current corresponds to a leftward propagation of
  $\delta\rho$ in the real space, from the liquid to the gas
  phase. Inspection of Fig.~\ref{FigCurrents} allows predicting all
  the dynamics reported in Fig.~\ref{PropagWaves}.
}\label{CorrespondenceDiagrams}
\end{figure}
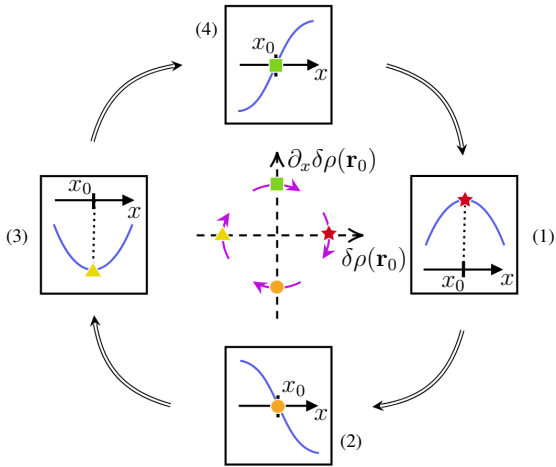

\textit{Propagating modes.} 
Let us now show how our formalism yields a valuable insight into the
dynamics of fluctuations. In Fig.~\ref{PropagWaves}a-c, we show the
short-time relaxations of a perturbation
$\delta\rho=\varepsilon\cos(qx)$ around a phase-separated profile
$\rho_s$ for $2\lambda+\kappa'$ negative, null or positive. To best
compare the three cases, we use the same $\rho_s$, corresponding to a
stationary droplet for $2\lambda+\kappa'=0$ (see
Fig.~\ref{PropagWaves}d). The analysis, detailed in
Fig.~\ref{CorrespondenceDiagrams}, of the current constructed in
Fig.~\ref{FigCurrents} predicts that the perturbation propagates at
the interface, from the liquid to the gas when $2\lambda+\kappa'<0$
and vice versa if $2\lambda+\kappa'>0$. In the equilibrium case, on
the contrary, the perturbation is predicted to relax to $\delta\rho=0$
while remaining stationary. These predictions are confirmed by the
simulations shown in Figs.~\ref{PropagWaves}a-c.

Figure~\ref{PropagWaves} shows the advection of initial perturbations
by the deterministic drift. In the presence of a finite noise,
superpositions of the corresponding propagating modes will be
constantly excited. Hence, to leader order in the noise, density
fluctuations propagate radially at the interface, outwards or inwards,
depending on the sign of $2\lambda+\kappa'$. This is the main
real-space manifestation of the steady-state probability current. It
suggests a natural mechanism to account for the continuous expulsion
of bubbles, from the bubbly liquid to the gas phase, observed in the
active model B +~\cite{tjhung2018cluster}. We note that higher order
contributions will also include orthoradial fluctuations. The latter
are particularly interesting since their dynamics could offer insight
on surface tension effects or capillary waves, which have recently
attracted a lot of
interest~\cite{bialke2015negative,marconi2016pressure,paliwal2017non,patch2018curvature,wittmann2019pressure,zakine2020surface,wysocki2020capillary,omar2020microscopic,lauersdorf2021phase}.

\textit{KPZ equation.}
Consider the celebrated KPZ equation~\cite{kardar1986dynamic}
\begin{equation}
\partial_t h = -\mu + \sqrt{2D}\Lambda \ , \quad \mu(\br,[h]) = \lambda |\nabla h|^2 -\kappa \Delta h \ .
\label{eq:KPZ}
\end{equation}
Here, $-\mu$ can be directly considered a vector field on $\mbF$
and the appropriate Riemannian metric to construct the vorticity is
the usual $L^2$-scalar product~\cite{supp}:
\begin{equation}
g_h(\mu_1,\mu_2) =\int \mu_1(\br,[h]) \mu_2(\br,[h]) \; \rmd \br \ .
\label{eq:KPZmetric}
\end{equation}
In this geometry, the one-form $\mu^\flat(\cdot) \equiv g(\mu,\cdot)$
associated to a vector field $\mu(\br,[h])$ again corresponds to
integration against $\mu$, and the vorticity of the deterministic
drift of Eq.~\eqref{eq:KPZ} is again $\bomega = -\mbd \mu^\flat$ where
$\mbd\mu^\flat=\int d \br (2 \lambda+\kappa') \nabla h \cdot
\delta_{\br}\wedge \nabla \delta_{\br}$. The analysis conducted above
for AMB can then be directly transposed to KPZ. We thus predict that
fluctuations around a height profile $h$ should again propagates
upward or downward $\nabla h$, depending on the sign of
$2\lambda+\kappa'$. These predictions are confirmed by the
simulations shown in Fig.~\ref{PropagWavesKPZ}.

\if{
Interestingly, the KPZ equation---that describes interface
growth---and AMB---that depicts nonequilibrium phase separation---show
significant differences but important similarities as well.  Indeed,
the deterministic drifts of these dynamics have different irrotational
components, hence leading to two different long term behaviors: while
the AMB dymanics converges to a fix point, the KPZ dynamics undergoes
unbounded growth. However, these deterministic drifts do share the
same vorticity, a fact that is made visible through similar
propagation of perturbations along gradients of the underlying field,
as shown in Fig.~\ref{PropagWaves}\&\ref{PropagWavesKPZ}.
}\fi

We stress that the KPZ equation and AMB describe fundamentally
different physics: the unbounded growth of a fluctuating interface and
the nonequilibrium phase separation of a conserved field leading to a
well defined stationary profile. These distinct long-term behaviors
stem from the different irrotational component of their determinist
drifts. Importantly, our analysis reveal that, on the contrary, these
systems share the same vorticity. In turn, this leads to an
unexpected similarity in the dynamics of fluctuations, as seen by
comparing Figs.~\ref{PropagWaves}\&\ref{PropagWavesKPZ}.

%


\textit{Discussion.}  In this Letter, we introduced a new formalism to
characterize the infinite-dimensional probability currents of
(stochastic) field theories without requiring the knowledge of their
stationary probability. This allowed us to determine their
low-dimensional measurable projections as well as to predict their
manifestations in the real, physical space. Our formalism is based
based on the generalization to functional spaces of the exterior
derivative. The latter offers a new, local and unambiguous criterion
to characterize the departure from equilibrium of coarse-grained
systems.

While we have focused here on AMB and the KPZ equation, both for sake
of clarity and due to the interest they have attracted over the years,
we note that our theoretical framework can be generalized to any
overdamped fluctuating hydrodynamics with Gaussian noise. Suppose for
instance that the chemical potential of Eq.~\eqref{SPDE}
or~\eqref{eq:KPZ} is given by a fourth order expansion in gradient of
$\rho$, i.e.  $\mu=\mu_0 + \lambda\vert\nabla\rho\vert^2 -
\kappa\Delta\rho + \alpha_1\Delta^2\rho +
\alpha_2\vert\nabla\rho\vert^4 +
\alpha_3\vert\nabla\rho\vert^2\Delta\rho + \alpha_4(\Delta\rho)^2 +
\alpha_5\nabla\rho\cdot\nabla\Delta\rho$, where $\mu_0$ is a local
function of $\rho$ and where, for the sake of clarity, all the other
coefficients are taken to be constant. This situation leads to
$\mathbb{d}\mu^\flat = \int d\br ( [2\lambda\nabla\rho +
  4\alpha_2\vert\nabla\rho\vert^2\nabla\rho +
  (\alpha_5-2\alpha_4)\nabla\Delta\rho]\cdot
\delta_\br\wedge\nabla\delta_\br +
\alpha_5\nabla\rho\cdot\delta_\br\wedge\nabla\Delta\delta_\br)$. Our
framework thus predicts the probability currents to be localized in
the spaces $(\rho(\br),\nabla\rho(\br))$ and
$(\rho(\br),\nabla\Delta\rho(\br))$. More generally, generalizing our
framework to non-local interactions, vector or tensor fields, as well
as for active mixtures~\cite{saha2020scalar,
  you2020nonreciprocity,dinelli2022self} is an exciting program for
the future.

\textit{Acknowledgments.} The author thanks Yariv Kafri, Julien Tailleur and Fr\'ed\'eric van Wijland for useful comments on the manuscript.

\bibliographystyle{apsrev4-1}
\bibliography{../../../Biblio}

\end{document}